\documentclass[a4,semcolor,psfig,english,12pt,epsf,portrait]{article}
\usepackage{amsmath,amsfonts,latexsym,amscd,amssymb,theorem}

\usepackage[T1]{fontenc}
\usepackage{epsfig}
\usepackage{amsmath}
\usepackage{psfrag}

\def\R{\hbox{\bf R}}
\def\Z{\hbox{\bf Z}}

\def\a{\alpha}

\def\e{\varepsilon}

\def\<{\langle}
\def\>{\rangle}
\newcommand{\ba}{\begin{eqnarray}}
\newcommand{\ea}{\end{eqnarray}}

\newtheorem{theo}{\bf Theorem}[section]
\newtheorem{lem}[theo]{\bf \textit{Lemma}}
\newtheorem{pro}[theo]{\bf Proposition}

\newtheorem{rem}[theo]{\bf Remark}

\renewcommand{\R}{{\mathbb R}}
\renewcommand{\Z}{{\mathbb Z}}

\begin{document}

\title{\bf Short time existence and uniqueness in \\ Hölder spaces  
for the 2D dynamics \\ of  dislocation densities }

\author{
\normalsize\textsc{ A. El
  Hajj$^1$}}
\vspace{20pt}
\maketitle
\footnotetext[1]{Universit\'e d'Orl\'eans,
Laboratoire MAPMO,
Route de Chartres, 45000 Orléans cedex 2, France.}


 \centerline{\small{\bf{Abstract}}}
 \noindent{\small{In this paper, we study the model of Groma and Balogh \cite{Groma}
     describing the dynamics of dislocation densities. This is a
     two-dimensional model where the
     dislocation densities satisfy a system of two  transport
     equations.  The velocity vector field is the shear stress in
     the material solving the equations of elasticity. This shear stress
     can be related to Riesz transforms of the dislocation
     densities.  Basing on some commutator estimates type, we show that
       this model has a unique local-in-time solution corresponding to any initial datum in the  space
     $C^r(\R^2)\cap L^p(\R^2)$ for $r>1$ and $1<p<+\infty$, where $C^r(\R^2)$ is the
     Hölder-Zygmund space.}}

\hfill\break
 \noindent{\small{\bf{AMS Classification: }}} {\small{54C70, 35L45, 35Q72, 74H20, 74H25.}}\hfill\break
  \noindent{\small{\bf{Key words: }}} {\small{Cauchy's problem,
      non-linear transport equations,  non-local transport 
equations,  system of hyperbolic equations, Riesz transform,
Hölder-Zygmund  space, dynamics of dislocation densities.}}\hfill\break


\section{Introduction}
\subsection{Physical motivation and presentation of the model}

Real crystals show certain  defects in the organization
of their crystalline structure, called dislocations. 
These defects were introduced in the Thirties by Taylor \cite{Tay},
Orowan \cite{Oro} and Polanyi  \cite{Pol} as the principal explanation of plastic
deformation  of materials at the microscopic scale. Dislocations can 
move under the action of exterior stresses applied to the material. \\

\noindent  Groma and  Balogh in \cite{Groma} considered the
 particular case where these defects are parallel lines in the three-dimensional
space, that can be viewed as points in a plane considering their cross-section.\\

\noindent In this model we consider two types of ``edge dislocations'' in
the plane $(x_1,x_2)$. Typically,  for a given velocity field,
those dislocations of type $(+)$ propagate in the direction  $+\vec{e_1}$ where $\vec{e_1}=(1,0)$
is the Burgers vector, while those of type $(-)$ propagate in the
direction $-\vec{e_1}$. We refer the reader to the book of   Hith and  Lothe \cite{Hirth}, for a
detailled description of the classical notion in physics of edge dislocations and of the 
Burgers vector associated to these dislocations.\\

\noindent In  \cite{Groma}  Groma and  Balogh have considered 
the case of  densities of dislocations. 
More precisely, this 2-D system
is given  by the following coupled non-local and non-linear 
transport equations (see Cannone et al.  \cite[Section 2]{EC} for
more modeling details):
 
\vspace{-0.8cm}
\begin{center}\begin{equation}\label{EC:eq:i:1}\left\{\begin{array}{lll}
\displaystyle{\frac{\partial{\rho^+}}{\partial t}(x,t)}+u
\displaystyle{\frac{\partial{\rho^+}}{\partial
    x_1}(x,t)}= 0 &\mbox{on $\R^2\times(0,T)$,}\\
 \\
  \displaystyle{\frac{\partial{\rho^-}}{\partial t}(x,t)}-u
\displaystyle{\frac{\partial{\rho^-}}{\partial
    x_1}(x,t)}= 0 &\mbox{on $\R^2\times(0,T)$,}\\
 \\
u= R_1^2R_2^2(\rho^+-\rho^-).
\end{array} \right.\end{equation}\end{center}

\noindent The unknowns of this system  are the two scalar functions $\rho^+$ and $\rho^-$
 at the time $t$ and the position $x=(x_1,x_2)$, that we denote for simplification by $\rho^{\pm}$. 
 This term correspond to the plastic deformations in a crystal. Its derivative in the  $x_1$-direction
$\displaystyle{\frac{\partial{\rho^{\pm}}}{\partial x_1}}$ represents
the dislocation densities of  type $(\pm)$.  Physically, these quantities
are non-negative. The function $u$ is the  velocity vector field which is equal to 
 the  shear  stress in the material, solving the equations of
 elasticity. The operators $R_1$ (resp. $R_2$)  are the
 $2D$ Riesz transform associated to $x_1$ (resp. $x_2$). More
 precisely,  the Fourier transform
 of these $2D$ Riesz transforms  $R_1$ and  $R_2$
are given by 

$$ \mbox{$\displaystyle{\widehat{R_k f}(\xi)=\frac
    {\xi_k}{|\xi|}\hat{f}(\xi)}$ \quad for  \quad  $\xi\in \R^2$, 
\quad $k=1,2$.}$$

\noindent  The goal of this work is to establish local existence and
uniqueness result of the solution of  (\ref{EC:eq:i:1})  
when the initial datum

\begin{equation}\label{EC:initial}{\rho^{\pm}}(x_1,x_2,t=0)
={\rho}^{\pm}_0(x_1,x_2)=\bar{\rho}^{\pm}_0(x_1,x_2)+Lx_1, \quad L\in \R
\end{equation}

\noindent  with  $\bar{\rho}^{\pm}_0 \in C^r(\R^2) \cap L^p(\R^2)$, for $r>1$, $p\in (1, +\infty)$,
where $C^r(\R^2)$ is the Hölder-Zygmund space defined in Section 2. The
choice $L>0$ guarantee the possibility to choose $\bar{\rho}^{\pm}_0\in
L^p(\R^2)$ such that the assumption is compatible with the non-negativity of 
$\displaystyle{\frac{\partial{\rho^{\pm}_0}}{\partial x_1}}$.  In a particular case where the  initial datum 
is  increasing, the global existence  of a solution was proved by  Cannone et al.
\cite{EC}, using especially an entropy inequality satisfies by the dislocation
densities. The fundamental issue of  uniqueness
for global solutions remains open. \\

\noindent In a particular sub-case of model (\ref{EC:eq:i:1}) where the
dislocation densities depend on a single variable $x_1+x_2$, the 
existence and uniqueness of a
Lipschitz solution was proved by El Hajj  et al. in 
\cite{EF} in the  framework of viscosity solutions. Also the existence
and uniqueness of a strong solution in $W^{1,2}_{loc}(\R\times
[0,+\infty))$ was proved by  El Hajj \cite{EL} in the  framework of
Sobolev spaces. For a similar model describing moreover boundary layer
effects (see  Groma,  Csikor,  Zaiser \cite{GromaZai}), we
refer the reader to Ibrahim \cite{Ibrahim} where a result  of
existence and uniqueness is established, using the framework of
viscosity solutions and also entropy solution for nonlinear hyperbolic equations. \\

\noindent Our study of the dynamics of dislocation
densities in a special geometry is related to the more general 
dynamics of dislocation lines. We refer the interested reader to the
work of  Alvarez et al.
\cite{AHLM04}, for a local
existence and uniqueness of some non-local Hamilton-Jacobi equation. We
also refer to  
Barles et al.  \cite{BCLM05}  for some long-time existence results.

\subsection{Main results}\label{EC:main}
We  shall show that the system (\ref{EC:eq:i:1}) possesses a unique
local-in-time solution for any 
initial datum satisfy  (\ref{EC:initial})  such that  
$\bar{\rho}^{\pm}_0  \in C^r(\R^2) \cap L^p(\R^2)$,  for  $r>1$ and for   $p\in (1,  +\infty)$. 
This  functional setting allows us to control the velocity
field $u$ in terms of $\rho^+ - \rho^-$ (see the third line of (\ref{EC:eq:i:1})). As we wrote it before, the velocity
 $u$ is related to $\rho^+ - \rho^-$ through the two-dimensional Riesz transforms $R_1$,
 $R_2$. Riesz transforms do not  map $C^r(\R^2)$ into $C^r(\R^2)$, but they are bounded on  $ C^r(\R^2)\cap L^p(\R^2)$, 
 for $r \in [0, +\infty)$ and for  $p\in (1,  +\infty)$, as we will see later.\\

\noindent For notational convenience,  we  define the space  $Y_ {r,p}$,
for $r\in  [0, +\infty)$ and $p\ge 1$  as follows

 $$Y_ {r,p}= \left \{\mbox  {$f=(f_1,f_2)$  such that   $f_k\in  C^r(\R^2) \cap L^p(\R^2)$,
  for $k=1,2$ }  \right\}, $$

\noindent where $C^r(\R^2)$ is the  inhomogeneous Hölder-Zygmund space (see
Section \ref{DEF}, 
 for more precise definition).  
This  space is  a Banach space  endowed with the following norm: for $f=(f_1,f_2)$ 

$$\left\|f\right\|_{r,p}= \max_{k=1,2}  \left (\left\|f_k\right\|_{C^r}
\right) +\max_{k=1,2} 
\left(\left\| f _k\right\|_{L^p} \right).$$

\noindent In order to avoid technical difficulties, we first consider
(see Theorem \ref{EC:theo:exi}) the case  $L=0$. Then
(see Theorem \ref{EC:theo:exi1}) we treat
the general case  $L \in \R$.

\begin{theo}\label{EC:theo:exi}{\bf (Local existence and  uniqueness, case $L=0$)}\\
Consider the initial data 

\begin{equation}\label{donnee_initiale}
{\rho}_0=({\rho}^{+}_0, {\rho}^{-}_0)\in Y_ {r,p}. 
\end{equation}
    
\noindent If   $r>1$ and $p\in (1, +\infty)$, then  
(\ref{EC:eq:i:1})
has a unique solution  $\rho=({\rho^{+}}, {\rho^{-}})\in L^{\infty}([0,T]; Y_ {r,p})$, where the time 
$T>0$ depends  only on $\left\|   \rho_0   \right\|_{r,p}$. Moreover,
the solution ${\rho}$ satisfies

$$ \rho \in Lip([0,T]; Y_ {r-1,p}).$$

\end{theo}

\noindent  In order to prove this theorem,  we strongly use  the fact that the
Riesz transforms  are  continuous on 
$C^r(\R^2)\cap L^p(\R^2) $ for $r\in [0, +\infty)$, $p\in (1, +\infty)$.  This result  ensures
that the velocity  vector
field remains  bounded on  $C^r(\R^2)\cap L^p(\R^2)$.  
Using this property and some commutator estimates, we can  prove that
 there exists some $T>0$  such that the solution $\rho^n$ of an approached
 system of (\ref{EC:eq:i:1}) (see system (\ref{4.5.EQ}) in Subsection 4.2)  is uniformly  bounded in
 $L^{\infty}([0,T]; Y_ {r,p})$ for $r > 1 $, $1<p < +\infty$.  Finally, 
 we show that the sequence of the  approximate solutions $\rho^n$ is a 
 Cauchy sequence in  $L^{\infty}([0,T]; Y_ {r-1,p})$, which  gives the
 local existence and uniqueness of the solution of (\ref{EC:eq:i:1}). \\

\noindent  The next theorem treats the general case  $L \in \R$.

\begin{theo}\label{EC:theo:exi1}{\bf (Local existence and  uniqueness, case $L \in \R$)}\\
Consider the equation (\ref{EC:eq:i:1}) corresponding to
initial data (\ref{EC:initial}), where $L\in \R$ and $\bar{\rho}_0=(\bar{\rho}^{+}_0,
\bar{\rho}^{-}_0)\in Y_ {r,p}$.  If   $r>1$ and $1<p < +\infty$, then  
(\ref{EC:eq:i:1})
has a unique solution  $\rho=({\rho^{+}}, {\rho^{-}}) \in L^{\infty}([0,T]; Y_
{r,p})$, where the time 
$T>0$ depends only  on  $L$ and $\left\| \bar{\rho}_0 \right\|_{r,p}$. Moreover,  

$${\rho}^{\pm}(x_1,x_2,t)= \bar{\rho}^{\pm}(x_1,x_2,t)+ Lx_1,$$ 

\noindent where 

$$\bar{\rho}=(\bar{\rho}^+,\bar{\rho}^-) \in  Lip([0,T]; Y_ {r-1,p}).$$

\end{theo}

\begin{rem}
If at the initial time we have $\displaystyle{\frac{\partial \rho^{\pm}}{\partial
    x_1}(\cdot,\cdot, t=0)}\ge 0$ two positive quantities, then this remains true
for $0 \le t\le T$, {\it i.e.}, 
$\displaystyle{\frac{\partial{\rho^{\pm}}}{\partial  x_1}} 
\ge 0$ for all $(x,t)\in \R\times [0,T]$. 
\end{rem}

\noindent Related to our analysis in the present paper, we get the
following theorem as a by-product.

\begin{theo}\label{EC:theo:exi2}{\bf (Global existence and  uniqueness for
  linear transport equations)}\\
Take $g_0 \in C^r(\R^2) \cap L^p(\R^2) $ and  $v=(v^1,v^2) \in L^{\infty}([0, T);  Y_ {r,p})$ for all $T>0$, $r>1$ and $1<p < +\infty$.
 Then, there exists a unique solution 

$$g \in L^{\infty}([0, T); C^r(\R^2)\cap L^p(\R^2))\cap Lip([0, T); C^{r-1}(\R^2)\cap L^p(\R^2))$$

\noindent of the linear transport equation 

\begin{equation}\label{transport}
\left \{ \begin{array}{ll}\displaystyle{\frac{\partial{g}}{\partial
 t}}+  v\cdot\nabla g=0 & \text{on}\quad \R^2\times(0,T) \\ \\
g(x, 0)=g_0(x) & \text{on}\quad\R^2.
\end{array}\right. \end{equation}

\end{theo}

\subsection{Organization of the paper}

This paper is organized as follows. In Section 2,
 we recall the characterization of  Hölder 
spaces and gather several important estimates. In particular, the boundedness of Riesz transforms 
 on $ C^r(\R^2)\cap L^p(\R^2) $ is established. Section 3 presents two key
 commutator estimates (Lemma \ref{3.1}). Finally in Section 4, we prove Theorem \ref{EC:theo:exi2} and a basic 
{\it a priori} estimate. Then, thanks to this  {\it a priori} estimate, we give 
in Subsections \ref{nphysique} and \ref{physique} the proofs of Theorems \ref{EC:theo:exi} 
and \ref{EC:theo:exi1} respectively.

\section{Some results on  Hölder-Zygmund spaces}\label{DEF}

This is a preparatory section in which we recall some results on 
 Hölder-Zygmund  spaces, and gather several  estimates that will be used
 in the subsequent sections. A
major part of the following results can be found in   Meyer \cite{Meyer1990} and 
Meyer, Coifman \cite{Meyer1991}.

 \noindent  We start with a dyadic decomposition of $\R^d$, where $d>0$
 is an integer. To this end, we take an arbitrary radial function  $\chi \in C_0^{\infty}(\R^d)$, such that
 
 $$  supp\; \chi \subset \left\{  \xi :  |\xi| \le \frac 43    \right\},
 \quad \chi \equiv 1\mbox{ for }  
 |\xi|  \le \frac 34,  
 \quad \|\chi\|_{L^1}=1.
  $$

 \noindent  It is a classical result that,  for $\phi(\xi)=\chi(\frac
 {\xi}{2})-\chi(\xi)$,  we have
 $\phi \in C_0^{\infty}(\R^d)$  and 
 
 $$ supp \;\phi \subset \left\{  
\xi :  \frac 34 \le  |\xi| \le \frac 83   \right \},$$
 
 $$ \chi(\xi)+ \sum_{j\ge 0} \phi(2^{-j}\xi)=1, \quad \mbox{for all  $ \xi\in \R^d$.}$$
 
  \noindent  For the purpose of isolating different Fourier frequencies,
  define the operators $\Delta_i$ for $i\in \Z$ 
as follows:
  
  \begin{equation}\label{2.1.EQ} \Delta_i f = \left \{\begin{array}{lc}
  0   &\mbox {if $i \le -2$},\\\\
  \chi(D)f= \displaystyle{\int \check{\chi}(y) f(x-y)dy} &\mbox {if $i=-1$},\\\\
  \phi(2^{-i} D)f= \displaystyle{2^{id}  \int \check{\phi}(2^{i}y) f(x-y)dy} &\mbox {if $i \ge 0$},
  \end{array}
  \right. \end{equation}
  
 \noindent   where  $\check{\chi}$ and  $\check{\phi}$ are the
 inverse Fourier transforms of $\chi$ and  $\phi$, 
respectively.
 
   \noindent For $i \in \Z$, $S_i$ is the sum of $\Delta_j$ with $j\le i-1$, i.e. 
   
   $$S_if= \Delta_{-1}f+\Delta_{0}f+\Delta_{1}f+ \cdots+ \Delta_{i-1}f=2^{id} \int \check{\chi}(2^{i}y) f(x-y)dy.$$

   \noindent It can be shown for any tempered distribution $f$ that
   $S_if \rightarrow f$ in the distributional sense, 
as $i \rightarrow \infty$. \\
   
  \noindent  For any $r \in \R$ and $p,q \in [1, \infty]$, the   inhomogeneous Besov
  space $B^r_{p,q}(\R^d)$ consists of all tempered
 distributions $f$ such that the sequence $\left \{2^{jr}
   \|\Delta_{j}f\|_{L^p} \right\}_{j\in \Z}$ belongs to $l^q(\Z)$. 
   When both $p$ and $q$ are equal to $\infty$, the Besov space
   $B^r_{p,q}(\R^d)$ reduces to the  inhomogeneous Hölder-Zygmund space $C^r(\R^d)$, i.e.
   $B^r_{\infty,\infty}(\R^d)=C^r(\R^d)$. More explicitly, $C^r(\R^d)$ with $r\in \R$
   contains any function $f$ satisfying
   
   \begin{equation}\label{CR}    \left\|f\right\|_{C^r}= \sup_{j\in \Z}
     2^{jr}    \|\Delta_{j}f\|_{L^{\infty}}< \infty. 
 \end{equation}
   
   \noindent It is easy to check that $C^r(\R^d)$ endowed with the norm
   defined in  (\ref{CR}) is a Banach space. \\

  \noindent  For $r\ge 0$, $C^r(\R^d)$ is closely related to the classical
  Hölder space  $\tilde{C}^r(\R^d)$ equipped with the 
norm

    \begin{equation}\label{CR1}  \left\|f\right\|_{\tilde{C}^r}=
      \sum_{|\beta| \le [r]} 
\left\|\partial^{\beta}f\right\|_{L^{\infty}}+ \sup_{x\neq y} \frac {
  \left |  \partial ^{[r]}f(x)-\partial ^{[r]}f(y)
    \right | }   {  \left |  x-y \right |^{r-[r]}}.  \end{equation}

  \noindent  In fact, if $r$ is not an integer, then the norm 
  (\ref{CR}) and  (\ref{CR1}) are 
equivalent, and ${C}^r(\R^d)=\tilde{C}^r(\R^d)$. The proof for this equivalence is classical and can be found in 
 Chemin \cite{Chemin}. When $r$ is an integer, say $r=k$,  $\tilde{C}^k(\R^d)$ is the
 space of bounded functions with 
bounded $j$-th derivatives for any $j\le k$. In particular,  
$\tilde{C}^1(\R^d)$ contains the usual Lipschitz functions and is sometimes
denoted by $Lip(\R^d)$. As a consequence of 
Bernstein's Lemma (stated below),  $\tilde{C}^r(\R^d)$ is a subspace 
of  ${C}^r(\R^d)$. Explicit examples can be constructed to show that such an
inclusion is genuine. In addition, according to 
Proposition \ref{2.2}, $\tilde{C}^r(\R^d)$ includes ${C}^{r+\e}(\R^d)$ for any $\e >0$. In 
summary, for any integer $k\ge 0$ and $\e >0$,

$${C}^{k+\e}(\R^d) \subset  \tilde{C}^k(\R^d) \subset   {C}^{k}(\R^d).$$

\begin{pro}\label{2.1}{\bf (Bernstein's Lemma},  Meyer \cite{Meyer1990} {\bf )}\\
Let $d>0$ be an integer and $\a_1 > \a_2 >0$ be two real numbers. 

  \noindent 1)  If $1\le p \le q \le \infty$ and $ supp \;  \hat{f}
  \subset \left 
\{  \xi \in \R^d:  |\xi| \le \a_1 2^j    \right \}$, then

$$\max_{|\a|=k}   \left\|\partial ^{\a}f\right\|_{L^{q}}  \le C
2^{jk+d(\frac 1p - \frac 1q)}  \left\|f\right\|_{L^{p}},$$

  \noindent  where $C>0$ is a constant depending only on $k$ and $\a_1$.

\noindent 2)   If $1\le p \le \infty$ and $ supp \;  \hat{f} \subset
\left \{  \xi \in \R^d:  \a_1 2^j 
 \le |\xi| \le \a_2 2^j    \right \}$, then 
$$ C^{-1} 2^{jk}  \left\|f\right\|_{L^{p}}  
\le 
\max_{|\a|=k}   \left\|\partial^{\a} f\right\|_{L^{p}}  \le C 2^{jk}  \left\|f\right\|_{L^{p}},   $$

  \noindent  where $C>0$ is a constant depending only on $k$, $\a_1$ and $\a_2$.

\end{pro}

\begin{pro}\label{2.2}{\bf (Logarithmic Sobolev inequalities in  Hölder-Zygmund
    space},    Kozono et al.
 \cite[Th 2.1]{Kozono} {\bf)}\\
Let $d>0$ be an integer. There exists a constant $C=C(d)$ such that for any $\e>0$ and $f \in
C^{\e}(\R^d)$, we have

\begin{equation}\label{2.4.EQ}  \left\|f\right\|_{L^{\infty}} \le \frac
  C\e \left\|f\right\|_{C^0} \log\left(e+ \frac
 {\left\|f\right\|_{C^\e}}{\left\|f\right\|_{C^0}}\right) \le 
\frac C\e \left\|f\right\|_{C^\e}. \end{equation}
\end{pro}

  \noindent  In the system (\ref{EC:eq:i:1}), the velocity field $u$
  is determined by
 $\rho^+-\rho^-$ through the $2D$ Riesz transforms. These Riesz transforms
  do not map a ${C^r}(\R^d)$  Hölder-Zygmund space to itself, but their
  action on $C^r(\R^d)$ is indeed bounded in $C^r(\R^d)\cap L^p(\R^d)$ for $p\in (1,
  +\infty)$ (see Proposition \ref{2.4}). We first recall a general result
 concerning the boudedness of Fourier multiplier operators on 
 Hölder spaces.

 \begin{pro}\label{2.3}{\bf (Fourier multiplier operators on 
 Hölder spaces},   Meyer \cite{Meyer1990}{\bf)}\\
 Let $d>0$ be an integer and $F$ be an infinitely differentiable function on $\R^d$. Assume that for
 some $R>0$ and $m \in \R$, we have
 
 $$ F(\lambda \xi)= \lambda^m F(\xi)$$
 
  \noindent   for any  $\xi \in \R^d$ with $|\xi|  > R$ and $
  \lambda \le 1$. Then the Fourier
 multiplier operator $F(D)$ maps 
 continuously  ${C^r}(\R^d)$ into $C^{r-m}(\R^d)$ for any  $r \in \R$.

 \end{pro}

 \begin{pro}\label{2.4}{\bf (Boundedness of Riesz transforms on $C^r(\R^d)\cap L^p(\R^d)$)}\\
 Let $r\in \R$ and  $p\in (1, +\infty)$. Then there exists a positive constant $C$ depending 
 only on $r$ and  $p$  such that

 $$\left\|R_kf\right\|_{C^r}  \le C  \left\|f\right\|_{C^r\cap L^p}  $$
  \noindent   where $k=1, 2$. 
 
  \end{pro}
 
\noindent {\bf Proof of Proposition \ref{2.4}:}\\
Using the operator $\Delta_{-1}$ defined in (\ref{2.1.EQ}), we divide $R_k f$  into two parts,

\begin{equation}\label{2.5} R_k f = \Delta_{-1}R_k f  + (1-\Delta_{-1}) R_k f .\end{equation}

\noindent Since $supp \; \chi(\xi) \cap supp \;
\phi(2^{-j}\xi)=\emptyset $ for $j \ge 1$, the operator 
$\Delta_{j}\Delta_{-1} = 0$ when 
 $j \ge 1$. Thus, according to (\ref{CR}),
 
 $$\begin{array}{lll}\left\|\Delta_{-1} R_kf\right\|_{C^r}
 &= \sup_{j\in \Z} 2^{jr} \left\|\Delta_{j} \Delta_{-1} R_kf\right\|_{L^{\infty}}\\\\
& =\max \left[  2^{-r} \left\|\Delta_{-1} \Delta_{-1}
    R_kf\right\|_{L^{\infty}},  \left\|\Delta_{0} \Delta_{-1}
 R_kf\right\|_{L^{\infty}} \right]\\\\
& \le \max \left[1, 2^{-r} \right]  \left\|\Delta_{-1} R_kf\right\|_{L^{\infty}}.
 \end{array}$$

\noindent  Let $q$ be the conjugate of $p$, namely $\frac 1p + \frac
1q=1$. It then follow, since
Riesz transforms are bounded on $L^p(\R^d)$, that for all
 $p\in (1, +\infty)$:
 
 $$ \begin{array}{lll}\left\|\Delta_{-1} R_kf\right\|_{C^r}
 &\le \max \left[1, 2^{-r} \right]  \left\| \check{\chi} \ast R_kf\right\|_{L^{\infty}}\\\\
  &\le \max \left[1, 2^{-r} \right]  \left \|\check{\chi} \right\|_{L^{q}} \left\|R_kf\right\|_{L^{p}}\\\\
&= C \left\| f\right\|_{L^{p}}
 \end{array}$$
 
 \noindent where $C$ is a constant depending only on $r$ and $p$. To
 estimate the second part in (\ref{2.5}), 
we apply Proposition \ref{2.3} 
 with  $\displaystyle{F(\xi)=(1-\chi(\xi))\frac {\xi_k}{|\xi|}}$ and
 $m=0$, and  hence we conclude that it maps 
${C^r}(\R^d)$ into $C^{r}(\R^d)$. This gives the proof of Proposition
\ref{2.4}. $\hfill\Box$ \\

  \noindent  Finally, we recall the notion of Bony's paraproduct (see 
 Bony \cite{Bony}). 
  The usual product $uv$ of two  functions $u$ and $v$ can be decomposed into 
  three parts. More precisely, using   $v= \displaystyle{\sum_{j\in \Z}
    \Delta_j v}$,  $u= \displaystyle{\sum_{j\in \Z}
 \Delta_j u}$  and
 
$$\mbox{  $\Delta_j\Delta_k v= 0$ if  $|j-k|\ge1$,  \quad
  $\Delta_j(S_{k-1} v \Delta_k v)= 0$ if  
$|j-k|\ge5 $, }$$ 
  
  \noindent    we can write 
  
  $$uv= T_u v+ T_v u + R(u,v),$$
  
   \noindent  where
  $$ T_u v =\sum_{j\ge 1} S_{j-1} (u)\Delta_j v,  \quad\quad R(u,v)=\sum_{|i-j|\le 1}  \Delta_i u \Delta_j v.$$

   \noindent  We remark that the previous decomposition allows one to distinguish different types of terms 
   in the product of $uv$. The Fourier frequencies of $u$ and $v$ in
   $T_u v$  and $T_v u$ are separated  
from each other while those 
   of the terms in $ R(u,v)$  are close to each other. Using this
   decomposition, one can show that 
   
  \begin{equation}\label{2.7}  \left\|uv\right\|_{C^s} \le  
 \left\|u\right\|_{C^s} \left\|v\right\|_{L^{\infty}}
 +\left\|u\right\|_{L^{\infty}} \left\|v\right\|_{C^s} \quad  \mbox{for} \quad s>0.  
\end{equation}
  
    \noindent  For the prove of  (\ref{2.7})  see    Chen \cite[Prop. 5.1]{Chen}.

\section{Two commutator estimates}

Two major commutator estimates are stated and proved in this section. We
remark that  this commutator estimates was often used to resolve the Navier-Stokes
equations (see for instance Cannone et al.  \cite{Cannone95},
\cite{Cannone2000}). Here, we apply these techniques on our system
(\ref{EC:eq:i:1}).

 \begin{lem}\label{3.1}{\bf ($L^{\infty}$ commutator estimates)}\\
 Let $j \ge -1$ be an integer and $r>0$. Then, for some absolute
 constant $C$, we have
 
 \noindent  1) 
 $$ \left \|\left[u \frac {\partial}{  \partial x_{\a}}, \Delta_{j}\right] f \right \|_{L^{\infty}} 
 \le C 2^{-jr} \left(    \left\| \frac {\partial f}{  \partial
       x_{\a}}\right\|_{L^{\infty}}  \left\|u\right\|_{C^r} 
  +     \left\| \nabla u\right\|_{L^{\infty}}
  \left\| f\right\|_{C^r}   
  \right),  \quad \mbox{for}  \quad \a=1,2,$$

  \noindent  2)  
  $$  \left \|\left[u \frac {\partial}{  \partial x_{\a}}, \Delta_{j}\right] f \right \|_{L^{\infty}} 
 \le C 2^{-jr} \left(    \left\| f \right\|_{L^{\infty}}
   \left\|u\right\|_{C^{r+1}}   +  
   \left\|{ \nabla u}\right\|_{L^{\infty}}
   \left\| f \right\|_{C^r}  
   \right), \quad \mbox{for} \quad  \a=1,2,$$
 
 \noindent where the bracket $[\; ,\;]$ represents the commutator, namely 
 
\begin{equation}\label{comm} \left[u \frac {\partial}{  \partial x_{\a}}, \Delta_{j}\right] f = u
 \frac {\partial}{
  \partial x_{\a}} (\Delta_{j} f)- \Delta_{j}\left(u\frac {\partial
    f}
{  \partial x_{\a}}\right), \quad \mbox{for} \quad  \a=1,2.
\end{equation}

 \end{lem}
\noindent {\bf Proof of Lemma \ref{3.1}:}\\
\noindent {\underline{ Proof of 1)}:}  Using the paraproduct notations $T$ and $R$, we decompose 
$\displaystyle{\left[u \frac {\partial}{  \partial x_{\a}}, \Delta_{j}\right] f}$, for  $\a=1,2,$  
into five parts,  

$$\left[u \frac {\partial}{  \partial x_{\a}}, \Delta_{j}\right] \rho= I_1+ I_2+I_3+ I_4+ I_5,$$

\noindent where \\

 \noindent  $\displaystyle{I_1= \left[T_u \frac {\partial}{  \partial x_{\a}}, \Delta_{j}\right] f= T_u \left(
 \frac {\partial}{
  \partial x_{\a}} (\Delta_{j} f) \right)- \Delta_{j}\left(T_u\frac {\partial
    f}
{  \partial x_{\a}}\right)}$,\\

\noindent  $\displaystyle{I_2=- \Delta_j T_{ \frac {\partial f}{  \partial x_{\a}} } u}$,

\noindent $\displaystyle{I_3= T_{ \frac {\partial( \Delta_j f)}{  \partial x_{\a}} } u}$,

\noindent $\displaystyle{I_4=   R\left (u, \frac {\partial (\Delta_j f)}{  \partial x_{\a}}\right)}$,\\

\noindent  $\displaystyle{I_5=-  \Delta_j  R\left(u, \frac {\partial f}{  \partial x_{\a}}\right)}$.\\

\noindent  Back to the definition of $T$, we can write 

\begin{equation}\label{3.6.EQ}\begin{array} {lll}I_1
&=\displaystyle{ \sum_{k\ge 1} S_{k-1} (u )\Delta_j \frac {\partial
    (\Delta_k f)}{  \partial x_{\a}} -
 \Delta_j \left( \sum_{k\ge 1} S_{k-1}( u) \Delta_k \frac {\partial f}{  \partial x_{\a}}         \right )}\\\\
&= \displaystyle{ \sum_{k\ge 1} \left[   S_{k-1} (u) \Delta_j \frac
    {\partial (\Delta_k f)}{  \partial x_{\a}}    
- \Delta_j     S_{k-1} (u)  \frac {\partial (\Delta_k  f)}{  \partial x_{\a}}    \right]. }  
\end{array} \end{equation}

\noindent  Since $\Delta_j \Delta_k = 0$ for $|j-k|>1$ and

$$ supp \left[ \widehat{  S_{k-1}(u)  \frac {\partial (\Delta_k  f)}{  \partial x_{\a}} } \right]
 \subset \left\{  \xi : \frac {1}{3} 2^{k-2}\le  |\xi| \le \frac {5}{3}2^{k+1} \right \},$$

\noindent the sum in (\ref{3.6.EQ}) only  involves those terms with $k$
satisfying $|j-k|\le 4$. We only take
 $j\ge 0$ since the case $j=-1$ can be handled similarly. Applying the definition of $\Delta_j$ 
in (\ref{2.1.EQ}), we obtain 

 $$ \begin{array} {lll} I_1
& =  \displaystyle{ \sum_{|j-k|\le 4}  2^{jd} \int \check{\phi}(2^j (x-y))  \left[
    S_{k-1}(u(x)) -  S_{k-1}(u(y)
) \right]  \frac {\partial (\Delta_k f)}{  \partial x_{\a}}(y) } dy\\\\
&= \displaystyle{   \sum_{|j-k|\le 4}  \int \check{\phi}(y)  \left[  S_{k-1}(u(x))
    -  S_{k-1}(u(x-2^{-j}y)) \right]  
\frac {\partial (\Delta_k  f)}{  \partial x_{\a}}(x-2^{-j}y) } dy.
\end{array} $$

\noindent Using  the fact that $\check{\phi} \in  S(\R^d)$   and  $S_j$  are continuous from
$L^{\infty}$ onto itself, we get 
 for $r\in \R$ and an absolute constant $C$:  

\begin{equation}\label{3.7.EQ}\begin{array} {lll}
\| I_1\|_{L^{\infty}}
 &\le C 2^{-j}  \| \nabla u \|_{L^{\infty}}   \left\|  \displaystyle{
     \frac {\partial (\Delta_j f)}{  
\partial x_{\a}} }\right \|_{L^{\infty}} \\\\
&\le  C  \| \nabla u \|_{L^{\infty}} \left\| \Delta_j f \right \|_{L^{\infty}} \\\\
&\le C 2^{-jr}   \| \nabla u \|_{L^{\infty}} \left\|  f  \right \|_{C^r},
 \end{array} \end{equation}
 
 \noindent  where we have used Proposition \ref{2.1} in the second inequality. 
 To estimate $I_2$ and $I_3$,  we first write them as 
  
  $$ I_2=-\displaystyle{ \sum_{|j-k|\le 4 }  \Delta_j \left( S_{k-1}
      \left (\frac {\partial f}{ 
 \partial x_{\a}}\right ) \Delta_k u\right)},  \quad \quad
  I_3= \displaystyle{ \sum_{|j-k|\le 4}  S_{k-1} \left (\frac {\partial
        (  \Delta_j  f)}{  \partial x_{\a}}\right)
 \Delta_k u}.$$

\noindent Similarly, only terms with $k$ satisfying $|j-k|\le 4$ are considered
in the above sums.  Thus,  since  
$\Delta_j$ and  $S_j$ 
 are continuous from $L^{\infty}$ onto itself, we have for $r \in \R$: 

\begin{equation}\label{3.81.EQ}
\| I_2\|_{L^{\infty}}
 \le C \| \Delta_j u \|_{L^{\infty}}   \left\|  \displaystyle{  S_{j}
     \frac {\partial f}{  \partial x_{\a}}}
\right \|_{L^{\infty}} 
 \le C 2^{-jr}  \|  u \|_{C^{r}}   \left\|  \displaystyle{\frac {\partial f}{  \partial x_{\a}}}\right \|_{L^{\infty}},
 \end{equation}

 \begin{equation}\label{3.82.EQ} \| I_3\|_{L^{\infty}}
 \le C \| \Delta_j u \|_{L^{\infty}}   \left\|  \displaystyle{  S_{j}
     \Delta_j  \frac {\partial f}{ 
 \partial x_{\a}}}\right \|_{L^{\infty}} 
 \le C 2^{-jr}  \|  u \|_{C^{r}}   \left\|  \displaystyle{\frac {\partial f}{  \partial x_{\a}}}\right \|_{L^{\infty}} ,
  \end{equation}

 \noindent   where the $C's$ in the above inequalities are absolute
 constants. From the definition of $R$, we have 
 
 $$ I_4= \displaystyle{ \sum_{|k_1-k_2| \le1}    \left(  \Delta_{k_1}
     (u)  \Delta_{k_2} \left(    
\displaystyle{\frac {\partial (\Delta_j f)}{  \partial x_{\a}}}  \right) \right) }.$$

  \noindent Obviously, only a finite number of terms involved in  the  above sums are non-zeros. Then,
  
 \begin{equation}\label{3.91.EQ}
\| I_4\|_{L^{\infty}}
 \le C \| \Delta_j u \|_{L^{\infty}}   \left\|  \displaystyle{
     \Delta_{j} \frac {\partial f}{  
\partial x_{\a}}}\right \|_{L^{\infty}} 
 \le C 2^{-jr}  \|  u \|_{C^{r}}   \left\|  \displaystyle{\frac
     {\partial f}{  
\partial x_{\a}}}\right \|_{L^{\infty}}.
 \end{equation}
 
   \noindent   Note that  from the definition of $R$ and $\Delta_j$,  $j \ge -1$, we can write $I_5$ as 
 
$$ I_5= -\displaystyle{ \sum_{k\ge j- 3 }\sum_{k_1 =  k- 1}^{k+1}
  \Delta_j  \left(  \Delta_{k} (u)  
\Delta_{k_1} \left(    \displaystyle{\frac {\partial f}{  \partial x_{\a}}}  \right) \right) }.$$

   \noindent    Therefore, for an absolute constant $C$, we have:

  \begin{equation}\label{3.92.EQ}\begin{array} {lll}
\| I_5\|_{L^{\infty}}
& \le C \displaystyle{ \sum_{k\ge j- 3 }\sum_{k_1 =  k- 1}^{k+1}  \|
  \Delta_k u \|_{L^{\infty}}   \left\|  \displaystyle{  
\Delta_{k_1} \frac {\partial f}{  \partial x_{\a}}}\right \|_{L^{\infty}}}\\\\
&  \le C \displaystyle{  \left\|  \displaystyle{\frac {\partial f}{
        \partial x_{\a}}}\right \|_{L^{\infty}}  
\|  u \|_{C^{r}}    \sum_{k\ge j- 3 }   2^{-kr} } \\\\
&  \le C\displaystyle{  2^{-jr}   \left\|  \displaystyle{\frac {\partial
        f}{  \partial x_{\a}}}\right 
\|_{L^{\infty}}  \|  u \|_{C^{r}} }. 
 \end{array}\end{equation}

   \noindent Gathering the estimates in (\ref{3.7.EQ})-(\ref{3.92.EQ}), we establish the desired inequality in 1).\\
   
\noindent {\underline{ Proof of 2)}:}    As in the proof of 1), we decompose 
 $\displaystyle{\left[u \frac {\partial}{  \partial x_{\a}}, \Delta_{j}\right] f}$ as the sum 
 of $I_1$, $I_2$, $I_3$, $I_4$ and $I_5$. The estimate on $I_1$ remains untouched, while different bounds are needed 
 for  $I_2$, $I_3$, $I_4$ and $I_5$.  Indeed, for $j \ge 1$: 

  \begin{equation}\label{3.10.EQ}\begin{array} {lll}
\| I_2\|_{L^{\infty}}
& \le \displaystyle{C \| \Delta_j u \|_{L^{\infty}}   \left\|
    \displaystyle{   \frac {\partial S_{j-1} f}{  
\partial x_{\a}}}\right \|_{L^{\infty}} }\\\\
& \le C 2^{j}  \|  \Delta_j u \|_{L^{\infty}}   \left\| S_{j-1} f \right \|_{L^{\infty}}\\\\
&\le C 2^{-jr}      \|  u \|_{C^{r+1}}     \left\| f \right \|_{L^{\infty}},
\end{array} \end{equation}

\noindent       where we have used Proposition \ref{2.1} in the second
inequality.  $I_3$ and $I_4$ can be similarly estimated as 
$I_2$: 
    \begin{equation}\label{3.11.EQ}\begin{array} {lll}
\| I_3\|_{L^{\infty}}
& \le \displaystyle{C    
\left\|  \displaystyle{   \frac {\partial \Delta_{j} f}{  \partial
      x_{\a}}}\right \|_{L^{\infty}} \| \Delta_j u \|_{L^{\infty}}}
\le C 2^{j}  \|  \Delta_j u \|_{L^{\infty}}   \left\| \Delta_j f \right \|_{L^{\infty}}\\\\
&\le C 2^{-jr}      \|  u \|_{C^{r+1}}     \left\| f \right
\|_{L^{\infty}},
\end{array} \end{equation}

    \begin{equation}\label{3.12.EQ}\begin{array} {lll}
\| I_4\|_{L^{\infty}}
& \le \displaystyle{C  \| \Delta_j u \|_{L^{\infty}}  
\left\|  \displaystyle{   \frac {\partial \Delta_{j} f}{  \partial
      x_{\a}}}\right \|_{L^{\infty}} }
 \le C 2^{j}  \|  \Delta_j u \|_{L^{\infty}}   \left\| \Delta_j f \right \|_{L^{\infty}}\\\\
&\le C 2^{-jr}      \|  u \|_{C^{r+1}}     \left\| f \right
\|_{L^{\infty}}.
\end{array} \end{equation} 

\noindent  Finally, we have
  \begin{equation}\label{3.13.EQ}\begin{array} {lll}
\| I_5\|_{L^{\infty}}
& \le C \displaystyle{ \sum_{k\ge j- 3 }\sum_{k_1 =  k- 1}^{k+1}  \|
  \Delta_k u \|_{L^{\infty}}  
 \left\|  \displaystyle{  \frac {\partial \Delta_{k_1} f}{  \partial
      x_{\a}}}\right \|_{L^{\infty}}}\\\\

&  \le C \displaystyle{ \sum_{k\ge j- 3 }\sum_{k_1 =  k- 1}^{k+1} 
2^{k_1} \|\Delta_k u \|_{L^{\infty}}  
 \left\|  \Delta_{k_1} f\right \|_{L^{\infty}}} \\\\

&   \le C \displaystyle{\left\| f\right \|_{L^{\infty} }
\sum_{k\ge j- 3 }
2^{k} \|\Delta_k u \|_{L^{\infty}}  
  } \\\\

& \le C \displaystyle{\left\| f\right \|_{L^{\infty}}\| u \|_{C^{r+1}}
\sum_{k\ge j- 3 }
2^{-kr}   
  } \\\\
& \le C \displaystyle{2^{-jr} \left\| f \right \|_{L^{\infty}} \| u \|_{C^{r+1}}.
  
 } 
 \end{array}\end{equation}
 
\noindent Combining  (\ref{3.10.EQ})-(\ref{3.13.EQ}) yields 2).

 $\hfill\Box$

\section{Local
existence and uniqueness results}

This section is devoted to the proofs of Theorems \ref{EC:theo:exi} and \ref{EC:theo:exi1}. For the sake of a clear presentation,
we divide it into three subsections. In the first subsection, we show a
basic {\it a priori} estimate and we prove Theorem \ref{EC:theo:exi2}. With the aid of this estimate, we prove 
 Theorems \ref{EC:theo:exi} and \ref{EC:theo:exi1} in the next subsections.

\subsection{An {\it a priori} estimate}

\begin{pro}\label{4.2}{\bf ({\it A priori} estimate)}\\
Let $r>1$ and $p>1$. For all  $T>0$, $\rho_0=(\rho_0^+, \rho_0^-) \in Y_{r,q}$ and 
$u\in L^{\infty}([0, T); C^r(\R^2) \cap L^p(\R^2))$, there exists a 
unique solution $\rho=(\rho^+, \rho^-)  \in L^{\infty}([0, T); Y_ {r,p})$ of the 
following system of linear transport equations 

\begin{equation}\label{linear}\displaystyle{\frac{\partial{\rho^\pm}}{\partial t}} \pm u
\displaystyle{\frac{\partial{\rho^\pm}}{\partial
    x_1}}= 0 \end{equation}

\noindent Moreover, for all $t \in [0,T]$, we have

$$ \left\|\rho (\cdot,t)\right\|_{r,p} \le  \left\|\rho_0 \right\|_{r,p}
exp\left( C \int_0^t 
  \left\| u (\cdot,\tau)\right\|_{C^r \cap L^p}  d\tau  \right)$$

\noindent   where $C>0$ is a constant depending only on $r$ and $p$.

\end{pro}

\noindent {\bf Proof of Proposition \ref{4.2}:}\\
From the fact that $u(\cdot,t) \in C^r(\R^2) \cap L^p(\R^2)$,  for $t\in [0,T]$, 
we can define the flow map $X^{\pm}(\cdot, t)$ satisfying

\vspace{-0.8cm}
\begin{center}\begin{equation}\label{4.2.EQ}\left\{\begin{array} {lll}
\displaystyle{\frac{\partial{X^{\pm}(x,t)}}{\partial t}}={\pm}\bar{u}(X(x,t),t), \quad \mbox{where}
\quad \bar{u}=(u,0), \\
 \\
X^{\pm}(x, 0)=x.
\end{array} \right.\end{equation}\end{center}

\noindent By the characteristics method, we know that, if 
$(X^\pm)^{-1}$ is the inverse function of $X^\pm$
with respect to $x$, then $\rho^{\pm}(x,t)=\rho^{\pm}_0( (X^\pm)^{-1}(x,t))$
is the unique solution of system  (\ref{linear})  (see   Serre \cite{Serre12} for more details). \\

\noindent  Let $j \ge -1$.  Applying  the operator $ \Delta_{j}$  to
both sides of the system (\ref{EC:eq:i:1})   yields

$$ \displaystyle{\frac{\partial{  \Delta_{j} \rho^{\pm}   }}{\partial t}}  {\pm} u
\displaystyle{\frac{\partial{ 
 \Delta_{j} \rho^{\pm}   }}{\partial x_1}} = {\pm}  \left[u \frac
{\partial}{  \partial x_1}, \Delta_{j}\right]
 \rho^{\pm},$$

\noindent  where $\displaystyle{\left[u \frac {\partial}{  \partial x_1},
  \Delta_{j}\right] \rho^{\pm}}$ is defined in (\ref{comm}). This equation can
be rewritten in the following form 

$$ \Delta_{j} \rho^{\pm}(x,t)=   \Delta_{j} \rho^{\pm}_0((X^{\pm})^{-1}(x,t)) {\pm} \int_0^t
\left[u \frac {\partial}{ 
 \partial x_1}, \Delta_{j}\right]   \rho^{\pm} ( X^{\pm}( (X^{\pm})^{-1}(x,t),s),  s) ds.$$

\noindent Taking the $L^{\infty}$-norm of both sides of this equality,
we get:

$$  \left\| \Delta_{j} \rho^{\pm}(\cdot,t) \right\|_{L^{\infty}} \le   \left\|
  \Delta_{j} \rho^{\pm}_0  \right\|_{L^{\infty}} 
 + \int_0^t  \left\| \left[u \frac {\partial}{  \partial x_1},
     \Delta_{j}\right] 
 \rho^{\pm}(\cdot,s) \right\|_{L^{\infty}}ds.$$

\noindent  Applying Lemma  \ref{3.1} (1), we obtain

$$  \left\| \rho^{\pm}(\cdot,t) \right\|_{C^r} \le  \left\|  \rho^{\pm}_0  \right\|_{C^r}  + C\int_0^t
\left(    \left\| \frac {\partial \rho^{\pm}   }{  \partial
      x_1}(\cdot,s)\right\|_{L^{\infty}}  \left\|u(\cdot,s)\right\|_{C^r}
   +     \left\| \nabla u (\cdot,s) \right\|_{L^{\infty}}  
  \left\|\rho^{\pm}(\cdot,s)\right\|_{C^r}     \right)
ds.$$

 \noindent  According to (\ref{2.4.EQ}), we know that for $r>1$ and a
 constant $C=C(r) >0$, we have:

$$  \left\|   \frac {\partial \rho^{\pm}  }{  \partial x_1}
\right\|_{L^{\infty}} \le  C \left\|
  \rho^{\pm} \right\|_{C^1} \log\left(e+ \frac {\left\| \rho^{\pm}
    \right\|_{C^r}}{\left\|
  \rho^{\pm} \right\|_{C^1}}\right) \le 
 C \left\|  \rho^{\pm} \right\|_{C^r}.  $$

\noindent  In a similar way, we can obtain $ \left\| \nabla u \right\|_{L^{\infty}}  \le C
\left\|  u \right\|_{C^r} $. 
Therefore, for $C=C(r) >0$,

$$ \begin{array}{ll}  \left\|  \rho^{\pm}(\cdot,t) \right\|_{C^r} 
& \displaystyle{ \le \left\| \rho^{\pm}_0
\right\|_{C^r} + C 
 \int_0^t  \left\| u(\cdot,s) \right\|_{C^r}  \left\| \rho^{\pm}(\cdot,s) \right\|_{C^r} ds}\\\\
  &   \displaystyle{\le \max_{\pm}\left(\left\| \rho^{\pm}_0
\right\|_{C^r}\right) + C 
 \int_0^t  \left\| u(\cdot,s) \right\|_{C^r}  \left\| \rho (\cdot,s) \right\|_{r,p} ds}
 .\end{array}$$

\noindent Moreover, integrating in time the  system  (\ref{EC:eq:i:1}), we get the following $L^p$  estimate:

$$\begin{array}{ll}  \left\| \rho^\pm (\cdot,t)  \right\|_{L^{p}}
& \displaystyle{\le  \left\|   \rho_0^\pm \right\|_{L^{p}} +  \int_0^t  \left\| u(\cdot,s) \right\|_{L^p}  \left\|    \frac {\partial \rho^{\pm}
}{  \partial x_1} (\cdot,s) \right\|_{L^{\infty}} ds}\\\\
& \displaystyle{ \le  \left\|  \rho_0^\pm \right\|_{L^{p}} + C \int_0^t  \left\| u(\cdot,s) \right\|_{L^p}  \left\|  \rho (\cdot,s) \right\|_{r,p } ds},
\end{array} $$

\noindent   where we have used  Hölder inequality in the first line  and  (\ref{2.4.EQ}) in the second line. Now,  adding 
the two previous inequalities, we obtain  

$$  \left\|  \rho(\cdot,t) \right\|_{r,p} \le \left\| \rho_0
\right\|_{r,p} + C 
 \int_0^t  \left\| u(\cdot,s) \right\|_{C^r \cap L^p}  \left\| \rho(\cdot,s) \right\|_{r,p} ds,$$

\noindent where $\|\cdot \|_{C^r \cap L^p}=\|\cdot \|_{C^r}+\|\cdot
\|_{L^p} $.  By Gronwall's Lemma and Proposition \ref{2.4},

\begin{equation}\label{EST_CR}\begin{array} {ll}\left\|\rho(\cdot,t)\right\|_{ r,p} 
  \displaystyle{\le  \left\|\rho_0\right\|_{r,p }
exp\left( C \int_0^t  
 \left\| u(\cdot,s)\right\|_{C^r \cap L^p}  ds  \right)}.
 \end {array}\end{equation}

\noindent Which completes the proof 
of Proposition \ref{4.2}.   

$\hfill\Box$

\noindent {\bf Proof of Theorem  \ref{EC:theo:exi2}:}\\
The proof of Theorem  \ref{EC:theo:exi2} is a consequence of the proof 
of Proposition \ref{4.2}. Indeed, just consider the characteristic 
equation

\vspace{-0.8cm} 
\begin{center}\begin{equation}\label{4.2.1.EQ}\left\{\begin{array} {lll}
\displaystyle{\frac{\partial{X(x,t)}}{\partial t}}=v(X(x,t),t), \\
 \\
X(x, 0)=x.
\end{array} \right.\end{equation}\end{center}

\noindent Then, as in the proof of Proposition \ref{4.2}, 
we use the commutator estimates proved in 
Lemma \ref{3.1} (1), to show the   following estimate:

\begin{equation}\label{EST_CR_L}\begin{array}{ll}\left\|g(\cdot,t)
\right\|_{C^r \cap L^p } 
  \displaystyle{\le  \left\|g_0\right\|_{C^r \cap L^p }
exp\left(C \int_0^t  
 \left\|v(\cdot,s)\right\|_{r,p} ds\right)},
 \end{array}\end{equation}

\noindent which proves the result.

$\hfill\Box$

\subsection{Proof of Theorem \ref{EC:theo:exi}}\label{nphysique}

The proof starts with the construction of a successive approximation
sequence $\{ \rho^n=(\rho^{+,n},\rho^{-,n} ) \}_{n \ge 1}$ satisfying
\vspace{-0.8cm}
\begin{center}\begin{equation}\label{4.5.EQ}\left\{\begin{array} {lll}
\rho^1= (\rho_0^+, \rho_0^-) =\rho_0,\\
 \\
 \displaystyle{\frac {\partial  \rho^{\pm, n+1} }{  \partial t} \pm u^n
   \frac {\partial  \rho^{\pm, n+1} }{  
\partial x_1}}=0, \quad \mbox{on} \quad \R^2 \times (0,T)\\\\
 u^n = R_1^2R_2^2\left(  \rho^{+, n}- \rho^{-, n}  \right), \\
 \\
 \rho^{\pm, n+1}(x,0)= \rho_0^\pm.
\end{array} \right.\end{equation}\end{center}

\noindent First of all, according to Proposition \ref{2.4}, 
$\rho^1 \in Y_{r,p}$ implies that $u^1 \in C^r(\R^2) \cap L^p(\R^2)$. Thus, applying Proposition 
\ref{4.2}, we can prove that, for all $T>0$,  there exists a unique solution 
$\rho^2 \in L^{\infty}([0,T); Y_{r,p})$ for (\ref{4.5.EQ}) with
$n=2$. Arguing in a similar manner  we can show that 
 this approached problem (\ref{4.5.EQ}) has a unique solution $\rho^n$ for
 all $n \ge 1$. \\

\noindent  The rest of the proof can be divided into two major steps. 
The first step establishes the
existence of $T_1 >0$ such that  $\{ \rho^n=(\rho^{+,n},\rho^{-,n})
\}_{n \ge 1}$ is uniformly  bounded  in $Y_{r,p}$
 for any $t \in [0,T_1]$.
The second step shows that for some $T_2 \in [0,T_1]$, we have   $\{ \rho^n
=(\rho^{+,n},\rho^{-,n}) \}_{n \ge 1}$  is a Cauchy sequence in 
$C( [0,T_2], Y_{r-1,p})$. \\

\noindent \underline{{\bf Step 1 (A uniform bound)}}: Using similar arguments as in the proof
of Proposition \ref{4.2}, estimate  (\ref{EST_CR}) yields, by Proposition \ref{2.4}, the
following bound on $\{ \rho^n=(\rho^{+,n},\rho^{-,n})\}_{n\ge 1}$:

$$\begin{array}{ll} \left\|\rho^{n+1}(\cdot,t)\right\|_{r,p} \le
& \displaystyle{\left\|\rho_0\right\|_{r,p}  exp\left( C_0 \int_0^t
   \left\|u(\cdot,s)\right\|_{C^r \cap L^p }  ds  \right)},\\
\\
& \displaystyle{\left\|\rho_0\right\|_{r,p}  exp\left( C_0 \int_0^t
   \left\|\rho^n(\cdot,s)\right\|_{r,p}  ds  \right)},
   \end{array}$$

\noindent where $r>1$,  $p\in (1, +\infty)$  and  $C_0=C_0(r,p)$. Choose $T_1$ and $M$ satisfying

$$M= 2\left\|\rho_0\right\|_{r,p}  \quad \mbox{and} \quad
  \left(\displaystyle{exp(C_0 T_1 M) \le 2} \quad \mbox{or}\quad \displaystyle{T_1=
 \frac{\ln( 2)}{2 C_0  \left\|\rho_0\right\|_{r,p}} }\right).$$

\noindent  Then $ \left\|\rho^n(\cdot,t)\right\|_{r,p} \le M$ for all $n
\ge 1$ and $t \in [0,T_1]$. Since,

$$\left\|\rho^1\right\|_{r,p}  \le  \left\|\rho_0\right\|_{r,p} <  M$$

\noindent and  $ \left\|\rho^k(\cdot,t)\right\|_{r,p} <  M$, we obtain 

\begin{equation}\label{EQ.LInfty}
  \left\|\rho^{n+1}(\cdot,t)\right\|_{r,p}  \le 
 \left\|\rho_0\right\|_{r,p}exp(C_0 T_1 M) \le M. \end{equation}

\noindent  Furthermore, since  $r>1$, we use (\ref{2.7}) and Proposition
\ref{2.4} to get:

$$ \begin{array} {lll} \left\|   \displaystyle{  \frac {\partial
        \rho^{\pm, n} }{ 
 \partial t}    }     \right\|_{C^{r-1}}
& \le    \left\|  \displaystyle{    u^n  \frac {\partial  \rho^{\pm,
        n+1} }{  
\partial x_1}     }  \right\|_{C^{r-1}} \\\\
&\le     \left\|u^n\right\|_{C^{r-1}} \left\| \displaystyle{ \frac
    {\partial  \rho^{\pm, n+1} }{  \partial x_1} }
 \right\|_{L^{\infty}} +\left\|u^n\right\|_{L^{\infty}} \left\|
   \displaystyle{ \frac {\partial  \rho^{\pm, n+1} }{ 
 \partial x_1}} \right\|_{C^{r-1}}\\\\
&\le   C \left\|   u^{n}   \right\|_{C^{r-1}}   \left\|\rho^{\pm, n+1}\right\|_{C^{r}}\\\\
&\le C M^2,
 \end{array}  $$
 
 \noindent  where $C=C(r)$.   We can also check that 
 the following $L^p$ estimate  on ${\rho}^{\pm, n}$ is valid:

 $$ \begin{array} {lll} \left\|   \displaystyle{  \frac {\partial   \rho^{\pm, n} }{  \partial t}  }     \right\|_{L^{p}}
& \le    \left\|  \displaystyle{    u^n  \frac {\partial \rho^{\pm, n+1} }{  \partial x_1}  }  \right\|_{L^{p}} 
\le     \left\|u^n\right\|_{L^{p}} \left\|  \displaystyle{    \frac {\partial \rho^{\pm, n+1} }{  \partial x_1}  } 
 \right\|_{L^{\infty}} \\\\
&\le   C \left\|   u^{n}   \right\|_{L^p}   \left\|   \rho^{\pm, n+1}\right\|_{C^{r}}
\le C M^2,
 \end{array}  $$
 
 \noindent  where we have used  Hölder inequality in the first line,
 then  (\ref{2.4.EQ}) and Proposition  \ref{2.4} in the second line.   
 Adding the two previous inequalities, we deduce that 
\begin{equation}\label{EQ.LIP}
 \left\|   \displaystyle{  \frac {\partial  \rho^{\pm, n} }{  \partial t}    }     \right\|_{C^{r-1}} 
+  \left\|   \displaystyle{  \frac {\partial  \rho^{\pm, n} }{  \partial t}    }     \right\|_{L^{p}}
\le C M^2,\end{equation}

 \noindent where $C=C(r)$.  Thus,   by
 (\ref{EQ.LInfty})-(\ref{EQ.LIP}), we obtain that 
 
 $$ \rho^{n} \in L^{\infty}([0,T_1]; Y_ {r,p})\cap   Lip([0,T_1]; Y_ {r-1,p} )$$

 \noindent is uniformly bounded. \\
 
 \noindent \underline{{\bf Step 2 (Cauchy sequence)}}: To show that  $\{ \rho^n =(\rho^{+,n},\rho^{-,n})  \}_{n \ge 1}$ is  a  Cauchy sequence in 
$ Y_{r-1,q}$, we consider the difference $\eta^{\pm, n}= \rho^{\pm, n} -\rho^{\pm, n-1}
$. Rigorously speaking, we should consider the more general difference
$\eta^{\pm, m,n}= \rho^{\pm, m} -\rho^{\pm, n}$, but the analysis for $\eta^{m,n}= (\eta^{+,m, n},\eta^{-,m,n})$ is
parallel to what we shall present for $\eta^{n}=(\eta^{+,n},\eta^{-,n})$
and thus we consider
$\eta^{n}$ for the sake of a concise presentation. It follows from
(\ref{4.5.EQ}) that $\{\eta^{n}=(\eta^{+,n},\eta^{-,n})\}_{n\ge 1}$ satisfies
\vspace{-0.8cm}
\begin{center}\begin{equation}\label{4.51.EQ}\left\{\begin{array} {lll}
\eta^{\pm,1}=\rho_0^\pm,\\
 \\
 \displaystyle{\frac {\partial  \eta^{\pm, n+1} }{  \partial t} \pm u^n  
\frac {\partial  \eta^{\pm, n+1} }{  \partial x_1}}= \mp w^n \frac {\partial
\rho^{\pm, n} }{  \partial x_1},
\\\\
 w^n = R_1^2R_2^2 (\eta^{+, n}-\eta^{-, n} ),\\
 \\
 \eta^{\pm, n+1}(x,0)=\eta^{\pm, n+1}_0(x)=0.
\end{array} \right.\end{equation}\end{center}
 
\noindent Proceeding as in the proof of Proposition \ref{4.2}, we obtain
for any integer $j\ge -1$,

 $$\begin{array} {lll}  \left\| \Delta_{j} \eta^{\pm, n+1}(\cdot,t) \right\|_{L^{\infty}}

&\displaystyle{\le \underbrace{\int_0^t  \left\| \left[u^{n} \frac {\partial}{  \partial x_1},
     \Delta_{j}\right] \eta^{\pm, n+1}(\cdot,s) \right\|_{L^{\infty}}ds }_{K_1}}
\\ \\
& + \displaystyle{  \underbrace{ \int_0^t  \left\|\Delta_{j}\left( w^n \frac {\partial
\rho^{\pm, n} }{  \partial x_1}(\cdot,s)
\right)\right\|_{L^{\infty}}ds. }_{K_2}}
\end{array}$$

\noindent Estimating $K_1$ by 
Lemma \ref{3.1} (2),  and $K_2$ by (\ref{2.7}), we get: 

 $$\begin{array} {lll}  \left\| \eta^{\pm, n+1}(\cdot,t)
   \right\|_{C^{r-1}} 
&\le \displaystyle{ C\int_0^t  
\left(  
   \left\|{ \nabla u^n(\cdot,s)}\right\|_{L^{\infty}}\left\|\eta^{\pm, n+1}(\cdot,s)\right\|_{C^{r-1}} + 
\left\|u^n(\cdot,s) \right\|_{C^{r}}\left\| \eta^{\pm, n+1}(\cdot,s) \right\|_{L^{\infty}} 
   \right)ds}\\ \\
&+\displaystyle{ C\int_0^t \left(\left\| w^n(\cdot,s) \right\|_{L^{\infty}}
   \left\|\frac {\partial\rho^{\pm, n} }{  \partial x_1}(\cdot,s) \right\|_{C^{r-1}}   +  
   \left\|w^n(\cdot,s)\right\|_{C^{r-1}}
   \left\|  \frac {\partial\rho^{\pm, n} }{  \partial x_1}(\cdot,s)   \right\|_{L^{\infty}}  
   \right)ds}.
\end{array}$$

\noindent  Since $r>1$,
Proposition \ref{2.2} implies, 

$$\|\nabla u^n\|_ {L^{\infty}} \le C\|u^n\|_ {C^{r}},\quad \quad
\|\eta^{\pm, n+1}\|_ {L^{\infty}} \le C\|\eta^{\pm, n+1}\|_ {C^{r-1}},$$

$$\left \| \frac {\partial\rho^{\pm, n} }{  \partial x_1} \right \|_ {L^{\infty}} \le
 C\|\rho^{\pm, n}\|_ {C^{r}} \quad \mbox{and} \quad
\|w^{n}\|_ {L^{\infty}} \le C\|w^{n}\|_ {C^{r-1}}.$$

\noindent Therefore, for a constant $C$  depending only on $r$,

$$\begin{array} {lll}  \left\| \eta^{\pm, n+1}(\cdot,t) \right\|_{C^{r-1}}
&\displaystyle{\le C\int_0^t\left\|{  u^n(\cdot,s)}\right\|_{C^{r}}  \left\|\eta^{\pm, n+1}(\cdot,s)\right\|_{C^{r-1}}
ds}\\ \\
& \;\;\; + \displaystyle{C\int_0^t \left\| w^n(\cdot,s) \right\|_{C^{r-1}}
   \left\|  \rho^{\pm, n}(\cdot,s) \right\|_{C^{r}}}ds.
\end{array}$$

\noindent However, it follows from a basic $L^p$ estimate that

$$\begin{array} {lll}  \left\| \eta^{\pm , n+1}(\cdot,t)\right\|_{L^{p }}
&\displaystyle{\le  
 C\int_0^t  \left\|{\nabla u^n(\cdot,s)}\right\|_{L^{\infty}}  \left\| \eta^{\pm, n+1} (\cdot,s)\right\|_{L^{p}}
ds}\\ \\
&\;\;\; + \displaystyle{C\int_0^t \left\| w^n(\cdot,s) \right\|_{L^{p}}
   \left \| \frac {\partial\rho^{\pm, n} }{\partial x_1}(\cdot,s) \right \|_ {L^{\infty}}}ds\\\\
&\displaystyle{\le  
 C\int_0^t  \left\|{ u^n(\cdot,s)}\right\|_{C^r}  \left\| \eta^{\pm, n+1} (\cdot,s)\right\|_{L^{p}}
ds}\\ \\
& \;\;\; + \displaystyle{C\int_0^t \left\| w^n(\cdot,s) \right\|_{L^{p}}
   \left\|  \rho^{\pm, n}(\cdot,s)\right\|_{C^{r}}}ds.
   
\end{array}$$

\noindent Adding the last two inequalities,  yields

$$\begin{array} {lll}  \left\| \eta^{n+1}(\cdot,t) \right\|_{r-1,p}
&\le   \displaystyle{ C\int_0^t \left\|{ u^n(\cdot,s)}\right\|_{C^{r} \cap L^p }\left\|\eta^{ n+1}(\cdot,s)\right\|_{r-1,p}}
ds\\ \\
&\;\;\; +\displaystyle{ C\int_0^t \left\| w^n(\cdot,s) \right\|_{C^{r-1} \cap L^p}
   \left\|  \rho^{ n} (\cdot,s) \right\|_{r,p }ds}.
   \end{array}$$

\noindent The components of $w^n$ are the Riesz transforms of $\eta^n$
and thus, according to Proposition \ref{2.4}: 
$$\|w^n\|_ { C^{r-1} \cap L^p } \le C\|\eta^n\|_ {r-1,p}.$$

\noindent We thus have reached an iterative relationship between
$\|\eta^n\|_ {r-1,p}$ and $\|\eta^{n+1}\|_ {r-1,p}$:

\begin{equation}\label{4.7.EQ}\begin{array} {lll}  \left\| \eta^{ n+1}(\cdot,t) \right\|_{r-1,p}
&\le  \displaystyle{ C_1\int_0^t \left\|  \rho^n(\cdot,s)\right\|_{r, p} \left\|\eta^{n+1}(\cdot,s)\right\|_{r-1,p}
ds}\\ \\
&\;\;\; +  \displaystyle{C_1\int_0^t \left\|\eta^n(\cdot,s) \right\|_{r-1,p}
   \left\| \rho^{ n} (\cdot,s) \right\|_{r,p }ds},
\end{array}\end{equation}

\noindent where the constants are labeled as $C_1$  for the purpose of
defining $T_2$. It has been shown in Step 1  that for $t\le T_1$,

$$ \|\rho^n\|_ {r,p} \le M.$$

\noindent Now, choose $T_2> 0$ satisfying

$$T_2 \le T_1,\quad \quad C_1 M T_2\le \frac 14,$$

\noindent  we can show that $\{\rho^n(\cdot,t)\}_{n\ge 1}$ is a Cauchy
sequence in $Y_{r-1,p}$ for $t\le T_2$. Indeed,
 for any given $\e >0$, if $\|\eta^{ n}\|_ {r-1,p}\le \e$ for $t\le T_2$, then
(\ref{4.7.EQ}) implies that:

$$\|\eta^{ n+1}\|_ {r-1,p}\le C_1\e
MT_2+C_1M\int_0^t\|\eta^{ n+1}(\cdot,s)\|_ {r-1,p}ds,$$

\noindent is valid for any $t\le T_2$. It then follows from Gronwall's
inequality that 

$$\|\eta^{n+1}\|_ {r-1,p} \le \e,$$

\noindent  for any $t\le T_2$ which completes Step 2.\\

\noindent  We conclude from Steps 1 and 2 that there exists $\rho=(\rho^+,\rho^-)$ satisfying 
 $$\rho \in L^{\infty}([0,T_2]; Y_ {r,p})\cap   Lip([0,T_2]; Y_ {r-1,p})$$

\noindent such that $\rho^n$ converges to $\rho$ in $C([0, T_2];  Y_ {r-1,p})$.

\noindent The proof of uniqueness follows directly from Step 2. 
This completes the proof of Theorem \ref{EC:theo:exi}.  $\hfill\Box$ 

\subsection{Proof of Theorem \ref{EC:theo:exi1}}\label{physique}
It is worth mentioning that the ideas of the proof of Theorem
\ref{EC:theo:exi1} are already contained in the proof of 
Theorem \ref{EC:theo:exi}.\\

\noindent First of all, we note that for all  $L\in \R$, if  $\rho^{\pm}$ are 
solutions of  (\ref{EC:eq:i:1}) then

$$\bar{\rho}^{\pm}(x_1,x_2,t)=\rho^{\pm}(x_1,x_2,t)- L x_1$$

\noindent  solves the following system:

\begin{equation}\label{linear1}\left\{\begin{array} {lll}
\displaystyle{\frac{\partial{\bar{\rho}^\pm}}{\partial t}(x,t)} \pm  u
\displaystyle{\frac{\partial{\bar{\rho}^\pm}}{\partial
    x_1}(x,t)}= \mp Lu  &\mbox{on $\R^2\times(0,T)$,}\\
 \\
u= R_1^2R_2^2(\bar{\rho}^+-\bar{\rho}^-),
\end{array} \right. \end{equation}

\noindent and respects the following initial data: 

$$\bar{\rho}^{\pm}_0(x_1,x_2,t)=\rho^{\pm}_0(x_1,x_2)- L x_1.$$

\noindent Now, to prove  Theorem \ref{EC:theo:exi1}, it suffices to
show that, for all initial data  $\bar{\rho}^{\pm}_0\in Y_{r,p}$, the
system (\ref{linear1}) has a unique local solution $\bar{\rho}^{\pm} \in
L^{\infty}([0,T); Y_{r,p})$ for  $r>1$ and $p \in (1, +\infty)$. \\

\noindent In order to do this, we proceed as in the proof of Theorem
\ref{EC:theo:exi}. We consider the following approached system:
\vspace{-0.8cm}
\begin{center}\begin{equation}\label{4.5.1.EQ}\left\{\begin{array} {lll}
\bar{\rho}^1=(\bar{\rho_0}^+, \bar{\rho_0}^-)=\bar{\rho}_0,\\
 \\
 \displaystyle{\frac {\partial  \bar{\rho}^{\pm, n+1} }{  \partial t} \pm u^n
   \frac {\partial  \bar{\rho}^{\pm, n+1} }{  
\partial x_1}}=\mp L u^n, \quad \mbox{on} \quad \R^2 \times (0,T)\\\\
 u^n = R_1^2R_2^2\left( \bar{\rho}^{+, n}-\bar{\rho}^{-, n}  \right), \\
 \\
 \bar{\rho}^{\pm, n+1}(x,0)= \bar{\rho_0}^\pm. 
\end{array} \right.\end{equation}\end{center}

\noindent We remark that, the only change that appears here, compared to
the approached system (\ref{4.5.EQ}) is the right-hand side $L u^n$ of
the second equation of (\ref{4.5.1.EQ}). However,
by Proposition \ref{2.4}, we know that this term remains bounded in 
 $L^{\infty}([0,T); C^r(\R^2) \cap L^p(\R^2))$ for $r
>1$ and $p \in (1, +\infty)$. Which  permits us to easily follow  the same
steps of the proof of Theorem \ref{EC:theo:exi}. This finally proves  that,
for some small $T>0$, we have on the one hand: the sequence  $\bar{\rho}^{n}= (\bar{\rho}^{+,
  n},\bar{\rho}^{-, n})$ is uniformly bounded in $L^{\infty}([0,T);
Y_{r,p})$, and on the other hand, this sequence is a Cauchy sequence in  $L^{\infty}([0,T);
Y_{r-1,p})$. This terminate the proof.

$\hfill\Box$



\section{Acknowledgements }
The author would like to thank M. Cannone and R. Monneau
 for fruitful remarks that helped in the preparation of the paper.
This work was partially
supported by  the program ``PPF, programme pluri-formations mathématiques
financières et EDP'', (2006-2010), Université Paris-Est.

 \bibliographystyle{siam}
 \bibliography{biblio}

\end{document}